# Pion Polarizability 2022 Status Report

## Murray Moinester

School of Physics and Astronomy
Tel Aviv University, 69978 Tel Aviv, Israel
E-mail: murray.moinester@gmail.com
http://murraymoinester.com

"Expanded version of talk to be presented at the 2022 APS Fall DNP Oct. 27 - 30, 2022 meeting in New Orleans, to be published as a review article."

**Abstract**

The electric $\alpha_\pi$ and magnetic $\beta_\pi$ charged pion Compton polarizabilities are of fundamental interest in the low-energy sector of quantum chromodynamics (QCD). They are directly linked to the phenomenon of spontaneously broken chiral symmetry within QCD and to the chiral QCD lagrangian. They characterize the induced dipole moments of the pion during $\gamma\pi$ Compton scattering. Pion polarizabilities affect the shape of the $\gamma\pi$ Compton scattering angular distribution. The combination $(\alpha_\pi - \beta_\pi)$ was measured by: (1) CERN COMPASS via $\pi^- Z \rightarrow \pi^- Z \gamma$ radiative pion Primakoff scattering (Bremsstrahlung) in the nuclear Coulomb field, (2) SLAC PEP Mark-II via two-photon production of pion pairs, $\gamma\gamma \rightarrow \pi^+\pi^-$, and (3) Mainz Microtron MAMI via radiative pion photoproduction from the proton, $\gamma p \rightarrow \gamma \pi^+ n$. Ongoing and planned pion polarizability experiments (COMPASS at CERN, BESIII at Beijing, JLab at Newport News) are also described. The Mark-II pion polarizability 95% confidence interval is approximately 0-11 $\times 10^{-4}$ fm$^3$, based on $\alpha_\pi - \beta_\pi = (4.4 \pm 3.2_{stat+syst}) \times 10^{-4}$ fm$^3$. Mainz's $\alpha_\pi - \beta_\pi = (11.6 \pm 1.5_{stat} \pm 3.0_{syst} \pm 0.5_{model}) \times 10^{-4}$ fm$^3$ is excluded on the basis of a dispersion relations calculation which uses the Mainz value as input, and gives significantly too large $\gamma\gamma \rightarrow \pi^0\pi^0$ sections compared to DESY Crystal Ball data. To date, only the COMPASS polarizability measurement has acceptably small uncertainties. Its value $\alpha_\pi - \beta_\pi = (4.0 \pm 1.8) \times 10^{-4}$ fm$^3$ agrees well with the two-loop ChPT prediction $\alpha_\pi - \beta_\pi = (5.7 \pm 1.0) \times 10^{-4}$ fm$^3$, thereby strengthening the identification of the pion with the Goldstone boson of QCD.

## 1. Introduction

Polarizabilities are well known via their association with the scattering cross section of sunlight photons on atomic electrons in atmospheric $N_2$ and $O_2$. At optical wavelengths, the incident photon energies are in the range 1.6 to 3.2 eV. Their oscillating (vector) electric field **E(t)** induces an oscillating (vector) electric atomic dipole moment **d(t)**; where **d(t)**=$\alpha_E$ **E(T)**, and the constant of proportionality $\alpha_E$ is known as the electric polarizability. The atom becomes a radiating dipole, whose radiation is seen as scattered light, that is alternatively described via a scattering cross section. Since this scattering cross section depends on $\lambda^{-4}$, the intensity of scattered and transmitted sunlight (through the atmosphere) is dominated by blue and red, respectively [Mo19]. Such scattering is denoted by Rayleigh scattering, following Lord Rayleigh's explanation of blue skies and red sunrises and sunsets [Ra99].

The electric $\alpha_\pi$ and magnetic $\beta_\pi$ charged pion polarizabilities characterize the induced dipole moments of the pion during $\gamma\pi$ Compton scattering. These moments are induced via the interaction of the $\gamma$'s electromagnetic field with the quark substructure of the pion. In particular, $\alpha_\pi$ is the proportionality constant between the $\gamma$'s electric field and the electric dipole moment, while $\beta_\pi$ is similarly related to the $\gamma$'s magnetic field and the induced magnetic dipole moment [Mo19, Mo98]. The polarizabilities are fundamental characteristics of the pion. A stringent test of chiral perturbation theory (ChPT) is possible by comparing the experimental polarizabilities with the chiral perturbation theory ChPT two-loop predictions $\alpha_\pi-\beta_\pi = (5.7\pm1.0)\times10^{-4}$ fm$^3$ and $\alpha_\pi+\beta_\pi = 0.16\times10^{-4}$ fm$^3$ [Ga06]. The pion polarizability combination ($\alpha_\pi-\beta_\pi$) may be measured by four different methods, as illustrated in Fig. 1. These are (1) radiative pion Primakoff scattering (pion Bremsstrahlung) in the nuclear Coulomb field $\pi Z \rightarrow \pi Z \gamma$, (2) two-photon fusion production of pion pairs $\gamma\gamma\rightarrow\pi\pi$ via the $e^+e^- \rightarrow e^+e^-\pi^+\pi^-$ reaction, (3) radiative pion photoproduction from the proton $\gamma p \rightarrow \gamma\pi^+ n$, and (4) Primakoff scattering of high energy $\gamma$'s in the nuclear Coulomb field leading to two-photon fusion production of pion pairs $\gamma\gamma\rightarrow\pi\pi$. Methods 1,2,3 experiments have been most recently studied at CERN COMPASS [Ad15], SLAC PEP Mark-II [Bo90], and Mainz Microtron MAMI [Ah05] respectively; while the method 4 experiment is planned at Jefferson Laboratory (JLab) [AL13]. The measured values are: (1) $\alpha_\pi-\beta_\pi = (4.0\pm1.2_{stat}\pm1.4_{syst})\times10^{-4}$fm$^3$, (2) $\alpha_\pi-\beta_\pi = (4.4\pm3.2_{stat+syst})\times10^{-4}$ fm$^3$,
(3) $\alpha_\pi-\beta_\pi = (11.6 \pm 1.5_{stat} \pm 3.0_{syst} \pm 0.5_{model}) \times10^{-4}$fm$^3$.

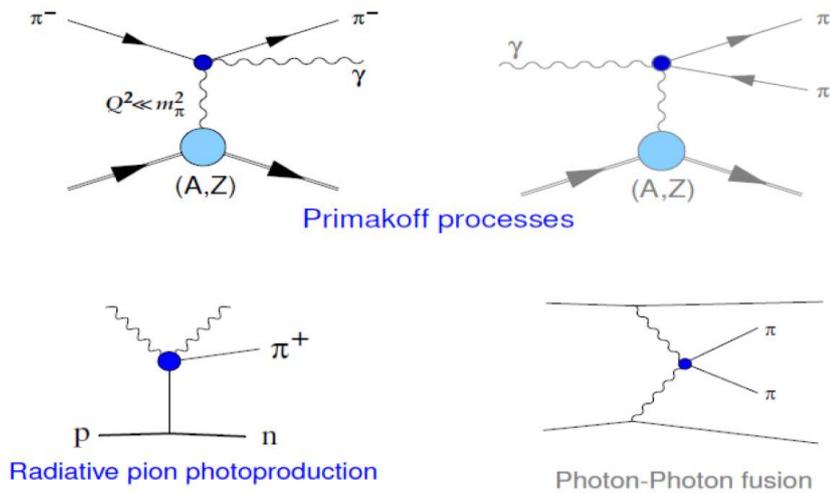

**Fig. 1:** Methods of studying pion polarizabilities: Primakoff Processes: Radiative pion scattering (Bremsstrahlung) on quasi-real photons in the nuclear Coulomb field; Primakoff scattering of high energy $\gamma$'s in the nuclear Coulomb field leading to two-photon fusion production of pion pairs, radiative pion photoproduction from the proton $\gamma p\rightarrow \gamma\pi n$; two-photon fusion production of pion pairs $\gamma\gamma\rightarrow\pi\pi$ via the $e^+e^- \rightarrow e^+e^-\pi^+\pi^-$ reaction.

A recent review of theory and experiment for the neutral pion polarizability is given in Section 5.3 of Ref. [Mo19], The theories are based on two-loop (and higher) ChPT and dispersion relations. The data are from DESY Crystal Ball [CB90], Belle [Bel08,Bel09], and BESIII [BES18, Bes19]. A future $\pi^0$ polarizability experiment is planned at Jefferson Laboratory [It20]. This subject will not be reviewed here.

## 2. COMPASS pion polarizabilities

CERN COMPASS investigates hadron-photon interactions to achieve a unique Primakoff Coulomb physics program, one of its many physics topics. The experiment scatters 190 GeV/c beams ($\pi$, K, p) from virtual photons in the Coulomb field of a target nucleus, and uses magnetic spectrometers and calorimeters to measure the complete kinematics of hadron-photon reactions. Thereby, pion polarizabilities, hybrid mesons, the chiral anomaly, and meson radiative transitions are studied.

COMPASS determined $\alpha_\pi - \beta_\pi$ by investigating pion Compton scattering $\gamma\pi \to \gamma\pi$ at center-of-mass energies below 3.5 pion masses [Ad15]. Compton scattering was measured via radiative pion Primakoff scattering (Bremsstrahlung of 190 GeV/c negative pions) in the nuclear Coulomb field of the Ni nucleus: $\pi^-$ Ni $\to \pi^-$ Ni $\gamma$. Exchanged quasi-real photons are selected by isolating the sharp Coulomb peak observed at lowest four-momentum transfers to the target nucleus, $Q^2 < 0.0015$ GeV$^2$/c$^2$. The resulting data are equivalent to $\gamma\pi \to \gamma\pi$ Compton scattering for laboratory $\gamma$'s having momenta of order 1 GeV/c incident on a target pion at rest. In the reference frame of this target pion, the cross section is sensitive to ($\alpha_\pi - \beta_\pi$) at backward angles of the scattered $\gamma$'s. This corresponds to the most forward angles in the laboratory frame for the highest energy Primakoff $\gamma$'s.

Pion Primakoff scattering at COMPASS is an ultra-peripheral reaction on a virtual photon target. The initial and final state pions are at a distance (impact parameter b) more than 100 fm from the target nucleus, significantly reducing meson exchange and final state interactions. This follows from the extremely small four-momentum transfer $Q_{min}$ to the target nucleus in a Primakoff reaction. For the COMPASS experiment, the minimum momentum transfer $Q_{min}$ ranges from 0.15 to 0.77 MeV/c; and the bulk of the cross section is in the range up to $3Q_{min}$. By the uncertainty principle, the average impact parameter $b \sim \hbar/(2Q_{min})$ is then larger than 100 fm.

COMPASS used a 190 GeV/c beam of negative hadrons (96.8% $\pi^-$, 2.4% $K^-$, 0.8% $\bar{p}$). The COMPASS spectrometer has a silicon tracker to measure precise meson scattering angles, electromagnetic calorimeters for $\gamma$ detection and for triggering, and Cherenkov threshold detectors for K/$\pi$ separation [Ab07]. Systematic uncertainties were controlled by many tests, including replacing pions by muons while keeping the same beam momentum. The muon Compton scattering cross section is precisely known, since muons have zero polarizabilities. From a 2009 data sample of 60,000 events, the extracted pion polarizabilities were determined.

Assuming $\alpha_\pi + \beta_\pi = 0$, the dependence of the laboratory differential cross section on $x_\gamma = E_\gamma/E_\pi$ is used to determine $\alpha_\pi$, where $x_\gamma$ is the fraction of the beam energy carried by the final state $\gamma$. The variable $x_\gamma$ is related to the $\gamma$ scattering angle for $\gamma\pi \to \gamma\pi$, so that the selected range in $x_\gamma$ corresponds to backward scattering, where the sensitivity to $\alpha_\pi - \beta_\pi$ is largest. Let $\sigma_E(x_\gamma)$ denote the experimental laboratory frame differential cross section as a function of $x_\gamma$. Furthermore, let $\sigma_{MC}(x_\gamma, \alpha_\pi)$ denote the calculated cross section for polarizability $\alpha_\pi$, using a Monte Carlo simulation, such that $\sigma_{MC}(x_\gamma, \alpha_\pi = 0)$ denotes the cross section for a point-like pion having zero polarizability. The $\sigma_E(x_\gamma)$ data are obtained after subtracting backgrounds from the
$\pi^-$ Ni $\to \pi^-$ Ni $\gamma$ diffractive channel and the $\pi^-$ Ni $\to \pi^-$ Ni $\pi^0$ diffractive and Primakoff channels. The ratios $R_\pi = \sigma_E(x_\gamma)/\sigma_{MC}(x_\gamma, \alpha_\pi = 0)$ are the experimental data points shown in Fig. 2. The polarizability $\alpha_\pi$ and its statistical error are extracted by fitting $R_\pi$ to the theoretical expression:

$$R_\pi = 1 - 72.73 \, \frac{x_\gamma^2}{1 - x_\gamma} \, \alpha_\pi,$$

where $\alpha_\pi$ is given in units of fm$^3$. The ratio $R_\pi$ with best fit $\alpha_\pi$ is shown in Fig. 2 as the solid curve [Ad15]. Systematic uncertainties were controlled by many tests, including measuring the $\mu^- Ni \to \mu^- Ni \gamma$ Primakoff cross sections by replacing pions by muons while keeping the same beam momentum. The muon Compton scattering cross section is precisely known, since muons have zero polarizabilities. The main contribution to the systematic uncertainties comes from the Monte Carlo description of the COMPASS setup. Comparing experimental and theoretical $x_\gamma$ dependences of $R_\pi$ yields: $\alpha_\pi = -\beta_\pi = (2.0\pm0.6_{stat}\pm0.7_{syst})\times10^{-4}$fm$^3$ or equivalently $\alpha_\pi-\beta_\pi = (4.0\pm1.2_{stat}\pm1.4_{syst})\times10^{-4}$fm$^3$, or $\alpha_\pi-\beta_\pi = (4.0\pm1.8)\times10^{-4}$fm$^3$. The COMPASS data analysis [Fr16, Gu10, Na12, Fr12] included corrections for one-photon-loop radiative effects [Ak95, Fr12], chiral loop effects [Gu10, Ka09], and the electromagnetic form factor of the Nickel nucleus. The corrections were modest, increasing the extracted polarizability values by $0.6\times10^{-4}$ fm$^3$ after they are applied [Ad15]. The ChPT one-loop calculation and dispersion relation calculations of Pasquini agree with one another on the permille level [Fr16].

Antipov et al. [An85] previously carried out a Primakoff polarizability experiment at Serpukhov using a 40 GeV/c beam of negative pions, and reported $\alpha_\pi-\beta_\pi = 13.6\pm2.8_{stat}\pm2.4_{syst} \times10^{-4}$fm$^3$, higher than the COMPASS result. However, since this low statistics experiment (~7000 events) did not allow complete precision studies of systematic errors, their result is not considered further in the present review.

**Fig. 2:** Determination of the pion polarizability by fitting the $x_\gamma$ distribution of the experimental ratios $R_\pi$ (data points) to the theoretical (Monte Carlo) ratio $R_\pi$ (solid line). (From Ref. 3)

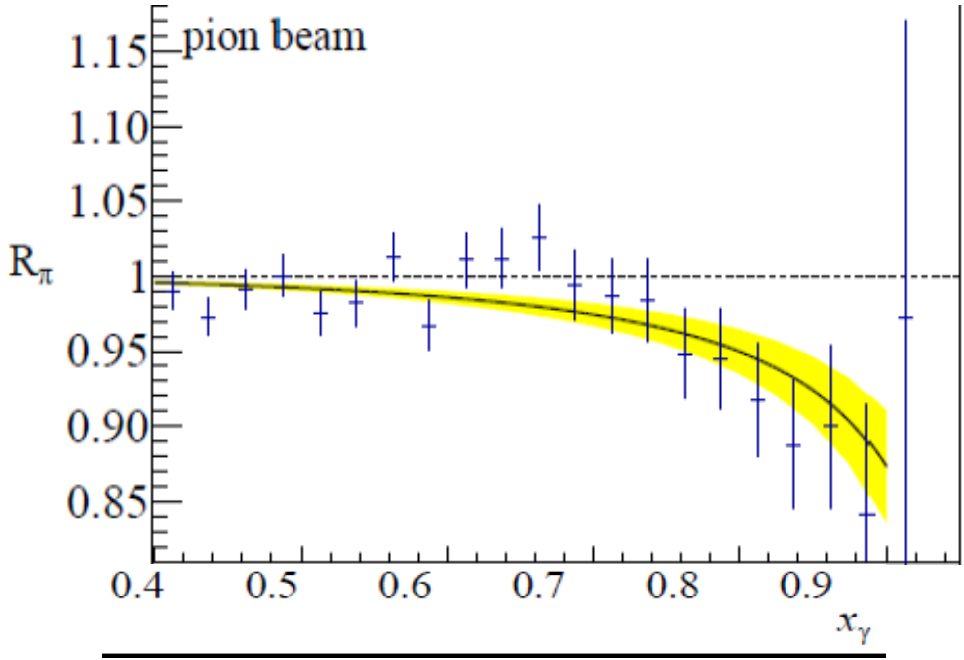

### 3. MAINZ pion polarizabilities

Radiative $\pi^+$-meson photoproduction from the proton ($\gamma p \to \gamma \pi^+ n$) was studied at the Mainz Microtron in the kinematic region 537 MeV $< E_\gamma <$ 817 MeV, $140° \leq \theta_{\gamma\gamma'} \leq 180°$, where $\theta_{\gamma\gamma'}$ is the polar angle in the c.m. system of the outgoing gamma and pion [Ah05]. The experimental challenge is that the incident $\gamma$-ray is scattered from an off-shell pion, and the polarizability contribution to the Compton cross section from the pion pole diagrams is only a small fraction of the measured cross section. The $\pi^+$-meson polarizability was determined from a comparison of the data with the predictions of two theoretical models, model-1 and model-2 [Ah05, Dr94]. Model-1 includes eleven pion and nucleon pole diagrams by using pseudoscalar coupling; while model-2 includes five nucleon and pion pole diagrams without the anomalous magnetic moments of the nucleons, diagrams for the $\Delta(1232)$, $P_{11}(1440)$, $D_{13}(1520)$, $S_{11}(1535)$ nucleon resonances, and σ-meson contributions. The validity of these two models was studied by comparing the predictions with the experimental data in the kinematic region where the pion polarizability contribution is negligible ($s_1 < 5m_\pi^2$), where $s_1$ is the square of the total energy in the $\gamma\pi^+ \to \gamma\pi^+$ c.m. system, and where the difference between the predictions of the two models does not exceed 3%. In the region where the pion polarizability contribution is substantial ($5 < s_1/m_\pi^2 < 15$; $-12 < t/m_\pi^2 < -2$), $\alpha_\pi - \beta_\pi$ was determined from a fit of the calculated cross section to the data, as illustrated in Fig. 3: $\alpha_\pi - \beta_\pi = (11.6 \pm 1.5_{stat} \pm 3.0_{syst} \pm 0.5_{model}) \times 10^{-4} \text{fm}^3$ [Ah05]. The quoted model uncertainty $0.5_{model} \times 10^{-4} \text{fm}^3$ denotes the uncertainty associated with using the two chosen theoretical models. It was estimated as half the difference between the model-1 and model-2 polarizability values. However, it does not take into account that comparisons with other possible models may significantly increase the model error. A larger model uncertainty could help explain the difference between COMPASS and Mainz polarizabilities.

It would be of interest to improve the estimate of the model uncertainty by using an independent model to extract the polarizability. A step towards a third model was taken by Kao, Norum, and Wang [Ka09] who studied the $\gamma p \to \gamma\pi^+ n$ reaction within the framework of heavy baryon chiral perturbation theory. They found that the contributions from two unknown low-energy constants in the πN chiral Lagrangian are comparable with the contributions of the charged pion polarizabilities. Their model therefore gives ~100% uncertainties for the existing Mainz data.

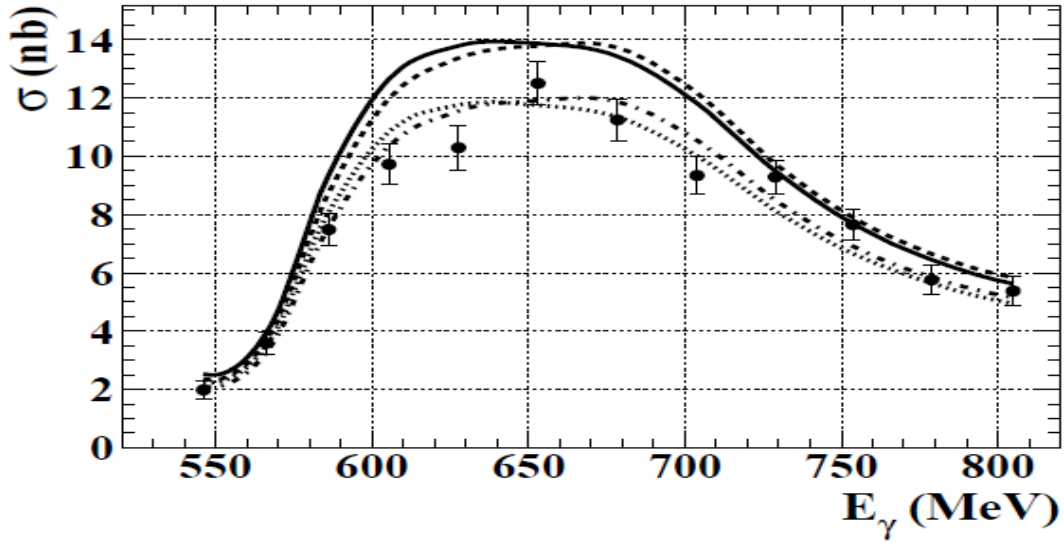

Fig. 3: The cross section of the process $\gamma p \to \gamma\pi^+ n$ integrated over $s_1$ and t in the region where the contribution of the pion polarizability is biggest and the difference between the predictions of the theoretical models under consideration does not exceed 3%. The dashed and dashed-dotted lines are predictions of model-1 and the solid and dotted lines of model-2 for $\alpha_\pi - \beta_\pi = 0$ and $14 \times 10^{-4} \text{fm}^3$, respectively [Ah05].

## 4. MARK-II pion polarizabilities

Charged pion polarizabilities were determined by comparing MARK-II total cross section data ($\gamma\gamma \to \pi^+\pi^-$) for $M_{\pi\pi} \leq 0.5$ GeV with a ChPT one-loop calculation [Ba92, Al94]. The MARK-II experiment was carried out via the reaction $e^+e^- \to e^+e^-\pi^+\pi^-$ at a center-of-mass energy of 29 GeV for invariant pion-pair masses $M_{\pi\pi}$ between 350 MeV/c$^2$ and 1.6 GeV/c$^2$ [Bo90]. Only the region below $M_{\pi\pi} = 0.5$ GeV is considered within the domain of validity of ChPT.

The most important problem in studying the $e^+e^- \to e^+e^-\pi^+\pi^-$ reaction is the elimination of the dominant two-prong QED reactions $e^+e^- \to e^+e^-e^+e^-$ and $e^+e^- \to e^+e^-\mu^+\mu^-$. These leptonic backgrounds below $M_{\pi\pi} = 0.5$ GeV are expected each to be more than 10 times larger than the expected signal. For the critical $M_{\pi\pi}$ region between 350 and 400 MeV/c, MARK-II eliminated these backgrounds by identifying pion pairs using time of flight (TOF), by requiring both tracks to hit an active region of the liquid-argon calorimeter, and by requiring both tracks to have a summed transverse momentum with respect to the $e^+e^-$ axis of less than 150 MeV/c [Bo90]. Summarizing, MARK-II at SLAC has the highest statistics and lowest systematic error data for $\gamma\gamma \to \pi^+\pi^-$ for $M_{\pi\pi} \leq 0.5$ GeV.

A number of theoretical papers subsequently made use of the MARK-II data to deduce pion polarizabilities. For example, theoretical curves from Refs. [Ba92, Al94] are shown in Fig. 4 for Born (dash-dotted line) and ChPT with $\alpha_\pi - \beta_\pi = 5.4 \times 10^{-4}$ fm$^3$ (full line). The cross section excess below $M_{\pi\pi} = 0.5$ GeV compared to the Born calculation was interpreted as due to pion polarizabilities, with best fit value $\alpha_\pi - \beta_\pi = (4.4 \pm 3.2_{stat+syst}) \times 10^{-4}$ fm. Similar analyses from Refs. [Ka94, Do93] gave $\alpha_\pi - \beta_\pi \sim 5.3 \times 10^{-4}$ fm$^3$, consistent with this result. The 95% confidence interval from these analyses is then approximately 0 to $11 \times 10^{-4}$ fm$^3$.

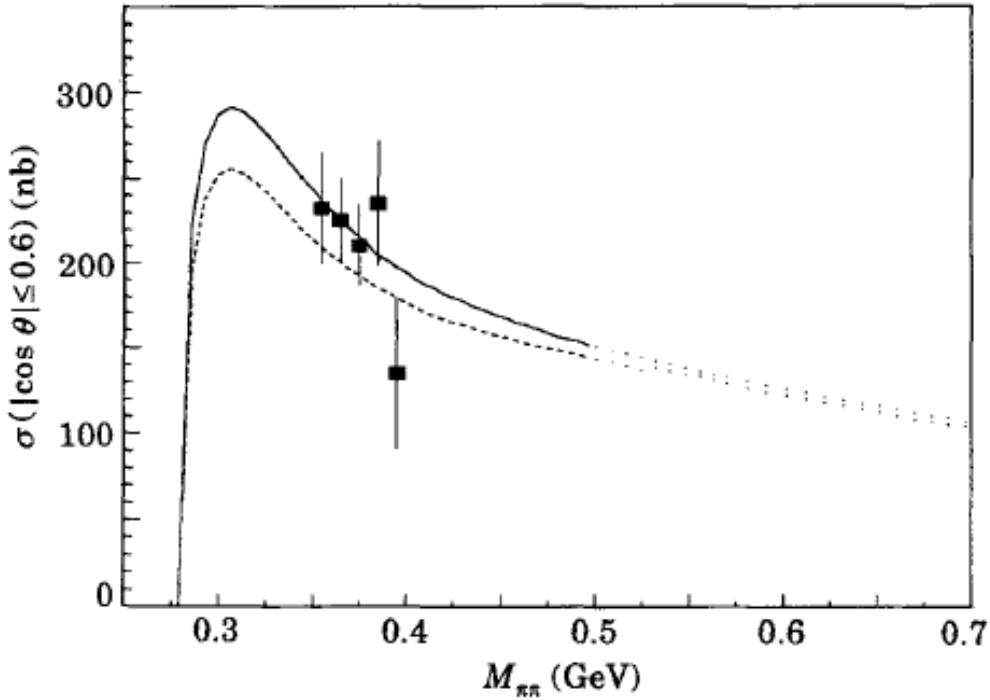

Fig. 4: MARK-II total cross section data ($\gamma\gamma \to \pi^+\pi^-$) for $M_{\pi\pi} \leq 0.5$ GeV. The theoretical curves are: Born (dash-dotted line); ChPT with $\alpha_\pi - \beta_\pi = 5.4 \times 10^{-4}$ fm$^3$ (full line). The region above $M_{\pi\pi} = 0.5$ GeV is considered outside the domain of validity of ChPT [Bo90].

## 5. Dispersion Relations and Pion Polarizabilities

Pion polarizabilities are determined by how the $\gamma\pi\to\gamma\pi$ Compton scattering amplitudes approach threshold. By crossing symmetry, the $\gamma\pi\to\gamma\pi$ amplitudes are related to the $\gamma\gamma\to\pi\pi$ amplitudes. Dispersion relations (DRs) provide the method to continue the $\gamma\gamma$ amplitudes analytically to the Compton scattering threshold. DRs describe how pion polarizabilities contribute to both $\gamma\gamma\to\pi^+\pi^-$, and $\gamma\gamma\to\pi^0\pi^0$ reactions [Be94, Bi98].

Most recently, Dai and Pennington (DP) carried out DR calculations [Da16]. In their formalism, the $\pi^0$ and $\pi^\pm$ polarizability values are correlated, so that knowing one allows calculating the other. Their calculation provides important guidance to the planned JLab $\gamma\gamma\to\pi^+\pi^-$ experiment [AL13], that the cross section must be measured to better than 2.2 nb to fix $\alpha_\pi-\beta_\pi$ to an accuracy of 10%. Using COMPASS $\alpha_\pi-\beta_\pi = 4.0\times10^{-4}$ fm$^3$ and Mainz $\alpha_\pi-\beta_\pi = 11.6\times10^{-4}$ fm$^3$ as input, DP calculate both $\gamma\gamma\to\pi^+\pi^-$, and $\gamma\gamma\to\pi^0\pi^0$ cross sections. They compare these with MARK-II $\gamma\gamma\to\pi^+\pi^-$ data and DESY Crystal Ball $\gamma\gamma\to\pi^0\pi^0$ data [CB90]. With the COMPASS value, they find excellent agreement for $\gamma\gamma\to\pi^+\pi^-$, and reasonable agreement for $\gamma\gamma\to\pi^0\pi^0$. With the Mainz value, their DR calculations and Crystal Ball data do not agree at all. The differences are too large to be explained by uncertainties in the DP calculation. DP conclude that $\alpha_\pi-\beta_\pi = 11.6\times10^{-4}$ fm$^3$ is excluded by the Crystal Ball $\gamma\gamma\to\pi^0\pi^0$ data. Following their evaluation, and also considering the large difference between Mainz and COMPASS values, the Mainz polarizability value is excluded from the Section 7 summary discussion.

Corrections for higher-order effects via a one-pion loop low energy expansion in ChPT increased the COMPASS polarizability values by $0.6\times10^{-4}$ fm$^3$ [Ad15]. Pasquini showed using subtracted Dispersion Relations (DR) for the pion Compton amplitude [Fr16, Pa17] that the yet higher order energy contributions neglected in the COMPASS analysis are very small. DRs take into account the full energy dependence, while ChPT uses a low energy expansion. In the COMPASS kinematic region of interest, the ChPT one-loop calculation and subtracted DRs agree in the mass range up to $4m_\pi$ at the two per mille level [FR16, Pa17]. Furthermore, the DR predictions of Pasquini, Drechsel, and Scherer [PA08, PA10], using unsubtracted DRs for the $\gamma\gamma\to\pi\pi$ amplitude, agree with the results of ChPT.

By contrast, Filkov & Kashevarov (FK) claim that there is significant disagreement between the ChPT one-loop calculation and their DR calculation [Fi05, Fi06, Fi10, Fi17, Fi18]. They claim therefore that the higher order corrections cannot be made via a one-pion loop calculation in ChPT. In their DR calculation, the contribution of the σ-meson to the COMPASS pion Compton scattering cross section is very substantial. FK claim that the COMPASS deduced ($\alpha_\pi-\beta_\pi$) is very sensitive to σ-meson contributions. They find via their DR calculations, taking into account the contribution of the σ-meson, that ($\alpha_\pi-\beta_\pi$) $\sim11\times10^{-4}$fm$^3$ for the COMPASS experiment.

Pasquini, Drechsel, and Scherer (PDS) claim however that the FK discrepancies arise due to the way that they implement the dispersion relations [Pa08, Pa10]. DRs may be based on specific forms for the absorptive part of the Compton amplitudes. The scalar σ-meson has low mass ($M_\sigma \sim 441$ MeV) and large width ($\Gamma_\sigma \sim 554$ MeV). In the DR calculation of FK, this resonance is modelled by an amplitude characterized by a pole, width and coupling constant. They choose an analytic form (small-width approximation, and an energy-dependent coupling constant) that leads to a dispersion integral that diverges like $1/\sqrt{t}$ for $t\to 0$. PDS examined the analytic properties of different analytic forms, and showed that the strong enhancement by the σ-meson, as found by FK, is connected with spurious (unphysical) singularities of this nonanalytic function. PDS explain that the FK resonance model gives large (unstable) results for the pion polarizability, because it diverges at t = 0, precisely where the polarizability is determined. That is, via DRs, the imaginary part of the Compton amplitudes serves as input to determine the polarizabilities at the Compton threshold (s = $m_\pi^2$, t = 0). PDS found that if the basic requirements of dispersion relations are taken into account, DR results and effective field theory are consistent. FK disagree with these PDS conclusions, claiming that their DR results are not due to spurious singularities. The present status report follows the views of Pasquini, Drechsel, and Scherer. Besides the σ-meson, modifications of COMPASS data analysis suggested by Filkov and Kashevarov [Fi10, Fi17] were studied, and found not to affect the deduced polarizabilities [Ad15].

### 6. Present and future experiments

Higher statistics data (~5 times) taken by COMPASS in 2012 is expected to provide an independent and high precision determination of $\alpha_\pi$ and $\beta_\pi$ without assuming $\alpha_\pi+\beta_\pi =0$ [Mo19, Fr16]. Measurement of the kaon polarizability would become possible if and when a radio-frequency-separated kaon beam becomes available.

Pion polarizability studies are also being carried out at the Beijing Spectrometer III (BESIII) at the Beijing Electron-Positron Collider II (BEPC II). BESIII is a cylindrically symmetric detector surrounding the interaction point of the $e^+e^-$ colliding beams. The experiment has collected high-statistics data samples for untagged and single-tagged events, $\gamma\gamma^* \to \pi^+\pi^-$ and $\gamma\gamma^* \to \pi^0\pi^0$ [Bes18, Bes19]. Currently, an analysis of single-tagged events is in an advanced state, for which the virtuality of one of the photons is in the range 0.1 GeV$^2$/c$^2$ to 4.0 GeV$^2$/c$^2$. QED $e^+e^-$ or $\mu^+\mu^-$ backgrounds are removed via high-quality particle identification and Monte Carlo simulations. Backgrounds with two pions in the final state are subtracted by fitting to the two-pion invariant mass spectrum.

The JLab pion polarizability experiment E12-13-008 [AL13], beginning ~2022, plans to measure $\gamma\gamma \to \pi^+\pi^-$ cross sections and asymmetries via the Primakoff_reaction $\gamma\gamma \to \pi^+\pi^-$. They will use a 6 GeV tagged linearly polarized photon beam produced via coherent bremsstrahlung, an Sn "virtual-photon target," the GlueX detector in Hall-D, and auxiliary detectors. GlueX is based on a solenoidal hermetic detector optimized for tracking of charged particles and detection of gamma rays. An important problem in studying the $\gamma\gamma \to \pi^+\pi^-$ reaction is the elimination of the two-prong QED reactions $\gamma\gamma \to e^+e^-$ and $\gamma\gamma \to \mu^+\mu^-$. The muon pair background below $M_{\pi^+\pi^-}$ = 0.5 GeV/c$^2$ is expected to be roughly 5 times larger than the expected signal; while the electron-positron background is expected to be negligible. TOF will not be useful for particle identification (PID) in the JLab measurement because of the extreme relativistic velocities of the pions. For $e^+/e^-$, they will use the GlueX FCAL lead-glass calorimeter along with tracking information to determine energy and momentum. For pion/muon PID, they will use a combination of the responses of the FCAL and two sets of downstream wire chambers separated by a passive iron hadronic absorber to distinguish pions from muons. The JLab experiment plans to use linearly polarized incident photons and to use the asymmetry and the azimuthal dependences of the $\pi^+\pi^-$ and $\mu^+\mu^-$ systems to help distinguish between signal and background. The advantages of measuring azimuthal correlations for both $\pi^+\pi^-$ and $\pi^0\pi^0$ channels was discussed in Refs. [On89, Al94]. Since the $\pi^+\pi^-$ channel will have contributions from both coherent $\rho^0$ photo-production on the nuclear target and Primakoff production of pion pairs, the asymmetry dependence will allow separating these channels. Dispersion relation calculations [Pa08, Da16] show that the total cross section for $|\cos(\theta)| < 0.6$ for $\gamma\gamma \to \pi^+\pi^-$ at $M_{\pi^+\pi^-} = 0.4$ GeV/c$^2$ equals approximately 170 nb and 210 nb for $\alpha_\pi-\beta_\pi$ equal to $5.7\times10^{-4}$ fm$^3$ and $13.0\times10^{-4}$ fm$^3$, respectively. These calculations provide important guidance in planning the JLab experiment that if $\alpha_\pi-\beta_\pi \sim 5.7\times10^{-4}$ fm$^3$, the cross section must be measured to about 3 nb uncertainty to achieve an accuracy of approximately 10%. Recent theoretical calculations for double pion photoproduction at forward angles and at threshold [Ge22] should provide important guidance for extracting pion polarizabilities from this planned JLab experiment.

### 7. Comparison of pion polarizability data with ChPT

The pion is believed to belong to the pseudoscalar meson nonet and to be one of the Goldstone bosons associated with spontaneously broken chiral symmetry. Chiral perturbation theory (ChPT) is therefore expected to successfully describe the electromagnetic interactions of pions. In this framework, the low-energy interactions of the pion are described by a phenomenological effective Lagrangian which stems directly from QCD, with only the assumptions of chiral symmetry SU(3)$_L$ X SU(3)$_R$, Lorentz invariance and low momentum transfer. Unitarity is achieved by adding pion loop corrections to lowest order, and the resulting infinite divergences are absorbed into physical (renormalized) coupling constants $L_i^r$ (tree level coefficients in the Lagrangian $L^{(4)}$, see ref. [Co77, Ru94]). In particular, with a perturbative expansion of $L^{(4)}$, limited to terms quartic in the external momenta and pion mass $O(p^4)$, the method establishes relationships between different processes in terms of a common set of renormalized parameters $L_i^r$. At $O(p^4)$ level, the perturbative expansion is truncated at terms quartic in the photon momentum and 12 coupling constants are needed. For example, in the charged pion case, the ratio of axial to vector form factor from radiative pion beta-decay and the electric polarizability are expressed as $h_A/h_V = 32\pi^2(L_9^r + L_{10}^r)$ and $\alpha_\pi = -\beta_\pi = 4\alpha(L_9^r + L_{10}^r)/m_\pi F_\pi^2$, where $F_\pi$ is the pion decay constant and $\alpha$ is the fine-structure constant [Co77, Ru94]. The electric $\alpha_\pi$ and magnetic $\beta_\pi$ charged pion Compton polarizabilities are therefore of fundamental interest in the low-energy sector of quantum chromodynamics (QCD). From the above description, they are directly linked to the phenomenon of spontaneously broken chiral symmetry within QCD and to the dynamics of the pion-photon interaction. The experimental pion beta-decay ratio
$h_A/h_V = 0.47 \pm 0.03$ [Wa50] then leads to the one-loop prediction $\alpha_\pi -\beta_\pi = 5.6\pm0.4\times10^{-4}$fm$^3$, assuming $\alpha_\pi+\beta_\pi=0$. The ChPT two-loop predictions are $\alpha_\pi-\beta_\pi = (5.7\pm1.0)\times10^{-4}$ fm$^3$ and $\alpha_\pi+\beta_\pi = 0.16\times10^{-4}$ fm$^3$ [Ga06]. The present report focuses on ChPT, and reviews all the different available pion polarizability experiments. Not within the framework of ChPT, Holstein [Ho90] and Xiong, et al. [Xo92] showed that meson exchange via a pole diagram involving the $a_1(1260)$ resonance provides the main contribution ($\alpha_\pi-\beta_\pi = 5.2\times10^{-4}$fm$^3$) to the polarizability. Other recent polarizability reviews are also available [AL16, Br16, Ec15, Fr16, Ho14, Ir16, La18, Mo19, Mo19B].

Pion polarizabilities affect the shape of the $\gamma\pi$ Compton scattering angular distribution. The pion polarizability combination ($\alpha_\pi-\beta_\pi$) was measured by: (1) CERN COMPASS via radiative pion Primakoff scattering (pion Bremsstrahlung) in the nuclear Coulomb field, $\pi Z \to \pi Z\gamma$, equivalent to $\gamma\pi \to \gamma\pi$ Compton scattering for laboratory $\gamma$'s having momenta of order 1 GeV/c incident on a target pion at rest; and (2) SLAC PEP Mark-II via two-photon production of pion pairs, $\gamma\gamma\to\pi^+\pi^-$. and (3) Mainz Microtron via radiative pion photoproductionfrom the proton, $\gamma p \to \gamma\pi^+ n$. To date, only the COMPASS polarizability measurement has acceptably small uncertainties. Its value $\alpha_\pi-\beta_\pi = (4.0\pm1.8)\times10^{-4}$ fm$^3$ agrees well with the two-loop ChPT prediction $\alpha_\pi-\beta_\pi = (5.7\pm1.0)\times10^{-4}$ fm$^3$, strengthening the identification of the pion with the Goldstone boson of QCD.

**Acknowledgments** Thanks are due to S. Scherer and B. Pasquini for helpful suggestions, to D. Lawrence for information concerning the planned JLab experiment, and to A. Denig, Y. Guo, and C. F. Redmer for discussions on the two-photon physics program at BESIII.

# 7. References

[Ab07] Abbon, P. et al. (2007). The COMPASS experiment at CERN. *Nucl. Instrum. and Meth. A577*, 455.

[Ad15] Adolph, C. et al. (2015). [COMPASS Collaboration], Measurement of the charged-pion polarizability, *Phys. Rev. Lett. 114,* 062002.

[Ah05] Ahrens, J. et al. (2005). Measurement of the $\pi^+$ meson polarizabilities via the $\gamma p \to \gamma \pi^+ n$ reaction. *Eur. Phys. J A23*, 113.

[Ak95] Akhundov, A. A., Gerzon, S., Kananov, S., Moinester, M. A. (1995). Radiative corrections for pion polarizability experiments, *Z. Phys. C 66,* 279.

[AL13] Aleksejevs, A., Miskimen, R., Lawrence, D., Smith, E. et al. [GlueX Collaboration] (2013). Measuring the Charged Pion Polarizability in the $\gamma\gamma \to \pi^+\pi^-$ Reaction, JLab PAC 40 approved Hall D proposal, *https://www.jlab.org/exp_prog/proposals/13/PR12-13-008.pdf*

[AL16] Aleksejevs, A., Barkanova, S. (2016). Electromagnetic Polarizabilities of Mesons, *Nucl. Part. Phys. Proc. 273*, 2773.

[Al94] Alexander, G. et al. (1994).Two-photon physics capabilities of KLOE at DAΦNE, *Nuovo Cimento 107A,* 837.

[An85] Antipov, Y. M. et al. (1985). Experimental estimation of the sum of pion electrical and magnetic polarizabilities, *Z. Phys. C26,* 495; Antipov, Y.M. et al. (1983). Measurement of $\pi^-$-meson polarizability in pion compton effect, *Phys. Lett. B121*, 445.

[Ba92] Babusci, D., Bellucci, S., Giordano, G., Matone, G., Sandorfi, A. M., Moinester, M. A. (1992). Chiral symmetry and pion polarizabilities, *Phys. Lett. B277*, 158.

[Be94] Bellucci, S., Gasser, J., Sainio, M. E. (1994). Low-energy photon-photon collisions to two-loop order, *Nucl. Phys. B423*, 80.

[Bel08] Uehara, S. et al. [Belle Collaboration] (2008). High-statistics measurement of neutral-pion pair production in two-photon collisions, *Phys. Rev. D78,* 052004.

[Bel09] Uehara, S. et al. [Belle Collaboration] (2009). High-statistics study of neutral-pion pair production in two-photon collisions, *Phys. Rev. D79,* 052009.

[Bes18] Redmer, C. F. [BESIII Collaboration] (2018). Measurements of Hadronic and Transition Form Factors at BESIII, *EPJ Web of Conferences* 212, 04004.

[Bes19] Guo, Y. [BESIII Collaboration] (2019). Two photon physics at BESIII, *J. Physics: Conference Series* 1137, 012008).

[Bi88] Bijnens, J., Cornet, F. (1988). Two-pion production in photon-photon collisions, *Nucl. Phys. B296*, 557.

[Bo90] Boyer, J. et al. (1990). Two-photon production of pion pairs, *Phys. Rev. D42,* 1350.

[Br16] Bressan, A. (2016). COMPASS Measurement of the Pion Polarizability, *Nucl. Phys. News 26,* 10.

[By09] Bychkov, M. et al. (2009). New Precise Measurement of the Pion Weak Form Factors in
$\pi^+ \to e^+ \nu \gamma$ Decay, *Phys. Rev. Lett. 103,* 051802.

[CB90] Marsiske, H. et al. [Crystal Ball Collaboration] (1990). Measurement of $\pi^0 \pi^0$ production in two-photon collisions, *Phys. Rev. D41*, 3324.

[Da16] Dai, L. Y., Pennington, M. R. (2016). Pion polarizabilities from a $\gamma\gamma \to \pi\pi$ analysis, *Phys. Rev. D94*, 116021.

[Do93] Donoghue, J. F., Holstein, B. R. (1993). Photon-photon scattering, pion polarizability, and chiral symmetry, *Phys. Rev. D48*, 137.

[Dr94] Drechsel, D., Filkov, L. V. (1994). Compton scattering on the pion and radiative pion photoproduction from the proton, *Z Phys. A349*, 177.

[EC15] Ecker, G. (2016). Status of chiral perturbation theory for light mesons, Chiral Dynamics Workshop, Proc. Sci. CD15, *https://pos.sissa.it/253/011/pdf*.

[Fi05] Filkov, L. V., Kashevarov, V. L. (2005). Determination of $\pi^0$ meson quadrupole polarizabilities from the process $\gamma\gamma \to \pi^0\pi^0$, *Phys. Rev. C72*, 035211.

[Fi06] Filkov, L. V., Kashevarov, V. L. (2006). Determination of $\pi^\pm$ meson polarizabilities from the
$\gamma\gamma \to \pi^+\pi^-$ process, *Phys. Rev. C73*, 035210.

[Fi10] Filkov L. V., Kashevarov V. L. (2010). Comment on "Polarizability of the pion: No conflict between dispersion theory and chiral perturbation theory", *Phys. Rev. C81*, 029801.

[Fi17] Filkov L.V., Kashevarov V.L. (2017).Dipole polarizabilities of charged pions, *Phys. Part. Nuclei 48*, 117.

[Fi18} Fi'kov, L. V., Kashevarov, V. L. (2018). *Int. J. Mod. Phys. Conf. Ser. 47*, 1860092.

[Fr12] Friedrich, J. M. (2012). *Chiral Dynamics in Pion-Photon Reactions: Habilitation* (Doctoral dissertation, Munich, Tech. U*.). https://inspirehep.net/files/ec968efd8050578ac47762bdbad642c5*

[Fr16] Friedrich, J. M. (2016). The pion polarisability and more measurements on chiral dynamics at COMPASS, Chiral Dynamics Workshop, Proc. Sci. CD15, *https://pos.sissa.it/253/015/pdf*

[Ga85] Gasser, J., Leutwyler, H. (1985). Chiral perturbation theory: expansions in the mass of the strange quark, *Nucl. Phys. B250*, 465.

[Ga06] Gasser, J., Ivanov, M. A, Sainio, M.E. (2006). Revisiting γγ→ π⁺π⁻ at low energies, *Nucl. Phys. B745*, 84

[Ge22] Gevorkyan, S., Larin, I., Miskimen, R., Smith, E. (2022). Photoproduction of pion pairs at high energy and small angles, *arXiv:2201.09923.*

[Gu10] Guskov, A. (2010). Analysis of the charged pion polarizability measurement method at COMPASS, *CERN-THESIS-2010-264.*

[Ho90] Holstein, B.R. (1990). Pion polarizability and chiral symmetry, *Comments Nucl. Part. Phys*. 19, 239.

[Ho14] Holstein, B. R., Scherer, S. (2014). Hadron polarizabilities, *Ann. Rev. Nucl. Part. Sci. 64*, 51

[It20] Ito, M.M., Zihlmann, B., Miskimen, R. et al. [GlueX Collaboration] (2020). Measuring the Neutral Pion Polarizability, JLab PAC 48 Hall D proposal. https://misportal.jlab.org/pacProposals/proposals/1538/attachments/127213/Proposal.pdf

[Iv16] Ivanov, M. A. (2016). Pion polarizabilities: ChPT vs Experiment. *EPJ Web of Conf. 129,* 00045. https://epjwoc.epj.org/articles/epjconf/pdf/2016/24/epjconf_qcd2016_00045.pdf

[Ka08] Kaiser, N., Friedrich, J. M. (2008). Cross-sections for low-energy πγ reactions, *Eur. Phys. J. A36*, 181.

[Ka94] Kaloshin, A. E., Serebryakov, V. V. (1994). π⁺ and π⁰ polarizabilities from γγ→ ππ data, *Z Phys. C64,* 689.

[Ka09] Kao, C. W., Norum, B. E., Wang, K. (2009). Extraction of the charged pion polarizabilities from radiative charged pion photoproduction in heavy baryon chiral perturbation theory, *Phys. Rev. D79*, 054001.

[La18] Lawrence, D. (2018). Measurements of Meson Polarizabilities, Proc. Sci. 317, **Chiral Dynamics Workshop,** *https://pos.sissa.it/317/032*

[Mo18] Moinester, M. (2018). Pion Polarizability Status Report, *arxiv:1709.05159*

[Mo19] Moinester, M., Scherer, S. (2019). Compton scattering off pions and electromagnetic polarizabilities, Int. J. Mod. Phys. A34, 1930008, *arxiv:1905.05640*

[Mo19B] Moinester, M. (2019). Pion Polarizability Review, Int. Conf. Quarks and Nuclear Physics, *JPS Conf. Proc. 26,* 021023, *https://journals.jps.jp/doi/pdf/10.7566/JPSCP.26.021023*

[Mo98] Moinester, M. A., Steiner, V. (1998). Pion and kaon polarizabilities and radiative transitions. *Chiral Dynamics: Theory and Experiment, Springer, 247, $arxiv:9801008$*

[Na12] Nagel, T. (2012). Measurement of the Charged Pion Polarizability at COMPASS, *Ph.D. thesis, TU Munich*, Section 5.2.

[On89] Ong, S., Kessler, P., Courau, A. (1989). Azimuthal correlations in double-tag measurements of photon-photon collisions, *Mod. Phys. Lett. A4*, 909.

[Pa08] Pasquini, B., Drechsel, D., Scherer, S. (2008). Polarizability of the pion: No conflict between dispersion theory and chiral perturbation theory, *Phys. Rev*. C77, 065211.

[Pa10] Pasquini, B., Drechsel, D., Scherer, S. (2010). Reply to "Comment on 'Polarizability of the pion: No conflict between dispersion theory and chiral perturbation theory". Phys. Rev. C81, 029802.

[Pa17] Pasquini, B. (2017). private communication.

[Ra99] Rayleigh, L. (1899). On the transmission of light through an atmosphere containing small particles in suspension, and on the origin of the blue of the sky, Lond. Edinb. Dublin philos. mag. j. sci. 47, 375.

[SC03] Scherer, S. (2003). Introduction to chiral perturbation theory, *Adv. Nucl. Phys. 27,* 277, *arxiv:0210398*

[Xo92] Xiong, L., Shuryak, E., Brown, G. (1992). Photon production through A1 resonance in high-energy heavy-ion collisions, Phys. Rev. 46D, 3798.